\documentstyle[12pt,aps,floats]{revtex} 

\begin{document}
\tighten
\draft
\title{
Quark Contributions to the Proton Spin and Tensor Charge
}

\vspace{0.3in}
\author{Han-xin  He
\footnotemark
\footnotetext{Email: hxhe@mipsa.ciae.ac.cn}
}
\vspace{0.1in}
\address{
China Institute of Atomic Energy \\ P. O. Box 275(18),
 Beijing 102413, China \\
and Institute of Theoretical Physics \\ Academia Sinica, Beijing 
100080, China 
}
\vspace{0.2in}
\date{AS-ITP-97XXX,~~~hep-ph/9712XXX,~~~October 1997}
\maketitle
\vspace{0.25in}
\begin{abstract}
    I calculate the quark contributions to the axial and tensor charges and the 
spin structure of the proton, which are related, respectively, through the 
transformations to their nonrelativistic quark model expectations. The 
result indicates that the valence 
current quark spins carry 1/3 of the proton spin, the total contribution of 
quark spins to the proton spin satisfies 
$\Delta \Sigma =1/3+\Delta \Sigma_{sea} \le 1/3$, and 
the quarks ( their spin plus orbital contributions ) contribute about one 
half of the proton spin at scale of 1 $GeV$. The valence current quark 
contributions to the proton tensor charge are also obtained.
\end{abstract}

\pacs{PACS numbers: 13.88.+e, 12.39.Jh, 14.20.Dh, 12.40.Aa}
\narrowtext
\vspace{0.35in}

The axial and tensor charges of the nucleon are the fundamental
observables charactering nucleon's properties. Measurement of these charges
leads to valuable insight into the spin-dependent quark substructure of the
nucleon because these charges are related through deep-inelastic sum rules
to the quark helicity distribution $\Delta q(x)$[1] and the transversity distribution 
$h_1(x)$[2], respectively. In recent years, of particular interest has been the 
flavor's singlet-axial charge which measures the contribution of quark 
spins to the nucleon spin. From recent data on polarized deep-inelastic
scattering(DIS)[3], one finds that about 30\% of the nucleon spin is carried
by quark spins[4], which deviates significantly from the expectation of the simple 
nonrelativistic constituent quark model(sNCQM). This discrepancy caused it to
be called "nucleon spin crisis".
As for nucleon's tensor charge[5,6], a first measurement of $h_1(x)$ (hence 
the tensor charge) will be undertaken at RHIC with transversely polarized 
Drell-Yan process and other processes[7]. Here one of important issues 
for understanding the proton spin puzzle and for predicting reliably the 
proton tensor charge is the relationship between the current (parton) quark 
and constituent quark contributions to the axial and tensor charges and 
to the spin structure of the nucleon[8], since in DIS and Drell-Yan processes
one probes the contribution from current quarks rather than constituent quarks. 
This Letter attempts to provide answers to them.

 In this Letter I present a quark model calculation of the quark contributions 
to the axial and tensor charges and the spin structure of the proton, which are related, 
respectively, through the transformations to their sNCQM's expressions. 
One thus shows that the valence constituent quark contributions to the axial 
and tensor charges and the spin structure of the proton in the traditional 
valence quark model include the contributions from the valence current quarks 
and the sea-($q\bar q$)-pairs which can be separated from each other. 
Antiquark polarizations, which can not be got through the trasformations 
hence must arise from the contributions beyond the traditional three quark
structure of nucleon such as the QCD anomaly etc, are 
introduced in the axial and tensor charges by definition and are found to be significant. 
Furthermore, I show a decomposition of the proton spin in the quark model, which 
is related through a transformation to the spin structure in sNCQM, and   
compare such a decomposition with the spin sum rule in QCD. 
 One finds that in the chiral limit of current quark mass $m_{cu}=0$ 
 the valence current quark spins carry $\frac 13$ of the proton 
spin, which is a basic origin of quark spin contribution measured in polarized 
DIS, suggesting that precise measurements for the quark spin contribution 
$\Delta \Sigma =1/3+\Delta \Sigma_{sea}$($\Delta \Sigma =0.337+\Delta \Sigma_{sea}$
if $m_{cu}\not=0$)
can lead to valuable insight into sea polarization inside the proton. 
The result of sum of the quark spin and 
orbital angular momentum indicates that quarks carry about one half the proton 
spin at scale of 1 $GeV$. The proton spin puzzle is interpreted. 

The nucleon axial and tensor charges can be written in terms of the nucleon
matrix elements of the axial quark current $\bar q \gamma^\mu\gamma^5 q$
and the tensor quark current $\bar q i\sigma^{\mu\nu}\gamma^5 q$, respectively:
\begin{equation}
\left\langle PS|\bar q\gamma ^\mu\gamma^5q|_{\mu^2}|PS\right\rangle 
=2\Delta q(\mu^2)S^\mu,
\end{equation}
\begin{equation}
\left\langle PS|\bar q i\sigma^{\mu\nu}\gamma ^5q|_{\mu^2}|PS\right\rangle 
=2\delta q(\mu^2)(S^\mu P^\nu-S^\nu P^\mu)
\end{equation}
with $q=u,d,s$, where $P$ and $S$ are the nucleon four-momentum and spin
vector, $\mu^2$ is a scale at which the operators are renormalized, or  
physically the nucleon wavefunction is probed.
$\Delta q$ and $\delta q$ are related to the quark distributions by
$\Delta q =\int_0^1dx[\Delta q(x)+\Delta \bar q(x)]
=\Delta q_{\it q}+\Delta \bar q$,
$\delta q=\int_0^1dx(h_1(x)-\bar h_1(x))=\delta q_{\it q}-\delta \bar q$.
Thus $\Delta q$ measures q-flavor contribution to the nucleon spin, and 
$\Delta q_{\it q}$ ($\Delta \bar q$) is longitudinal quark (antiquark)
polarization in longitudinal polarized nucleon, while
$\delta q_{\it q} $ ($\delta \bar q$) is transverse quark (antiquark) 
polarization in transverse polarized nucleon. If one
separates the quarks (partons) into valence current quarks and sea quarks and 
assumes that antiquarks are all in the sea[10], one may write $\Delta q$ and
$\delta q$ as: $\Delta q=\Delta q_{\it q}+\Delta \bar q=\Delta q_v+\Delta q_s
+\Delta \bar q=\Delta q_v+\Delta q_{sea}$, and $\delta q=\delta q_{\it q}-\delta
\bar q=\delta q_v+\delta q_s-\delta \bar q=\delta q_v$. The sea quarks do
not contribute to $\delta q$ because the tensor current operator is odd
under charge conjugate. Therefore, $\delta q $ only counts the valence current
quarks of opposite transversity.

   One now calculates $\Delta q$ and $\delta q$ based on Eqs.(1) and (2) by 
using the quark model. In the rest frame of the nucleon, Eqs.(1) and (2) become
$\langle PS|\bar q\gamma ^i\gamma ^5q|_{\mu^2}|PS\rangle 
=2\Delta q(\mu^2)S^i$ and
$\langle PS|\bar q\gamma ^0\gamma ^i\gamma ^5q|_{\mu^2}|PS\rangle =
2\delta q(\mu^2)S^i $, where $\mu^2$ takes 1 $GeV^2$,
 the scale that the constituent quark picture works. One notices that in 
these matrix elements the nucleon wavefunction in the quark model can be 
written in terms of the Dirac spinor for the quarks, while in the NCQM the 
nucleon wavefunction is traditionally written in terms of the Pauli spinor for 
the quarks. Thus, by transforming identically the Dirac spinor into the Pauli 
spinor and some calculations, I find that the axial charge $\Delta q$
and the tensor charge $\delta q$ defined by these matrix elements are 
related to their sNCQM expressions, 
$\Delta q_{NR}$ and $\delta q_{NR}$, as follows
\begin{equation}
\Delta q_{\it q}=\left\langle M_A\right\rangle \Delta q_{NR},
\end{equation}
\begin{equation}
\delta q_{\it q}=\left\langle M_T\right\rangle \delta q_{NR}
\end{equation}
with
\begin{equation}
M_A=\frac 13+\frac{2m}{3E},\ \ \ \ M_T=\frac 23+\frac m{3E},
\end{equation}
and
\begin{equation}
2 \delta q_{\it q}=\Delta q_{\it q}+\Delta q_{NR},
\end{equation}
where $\Delta q_{NR}=\langle P\uparrow |\chi_s^{+}\sigma_3^{(q)}\chi_s|P\uparrow \rangle$,     
$m$ and $E=\sqrt {m^2+\vec k^2}$ are quark mass and energy, respectively, and
$\langle M_{A,T}\rangle $ are the expectation values of
transformation matrices $M_{A,T}$ in the nucleon state. In the
sNCQM, the transverse polarized quarks are in the
transverse spin eigenstates, which by rotational invariance implies
 $\delta q_{NR}=\Delta q_{NR}$. For the proton, one has $\Delta u_{NR}
=4/3$, $\Delta d_{NR}=-1/3$, $\Delta s_{NR}=0$, $\Delta \Sigma_{NR}=
\Delta u_{NR}+\Delta d_{NR}+\Delta s_{NR}=1$, which expresses that the proton
spin is carried by the spins of the three static constituent quarks.

The derivation of Eqs.(3)-(6) is sketched below. One expands the quark field
operators in the nucleon matrix elements of quark currents in terms of a 
complete set of quark and antiquark wavefunctions. The $\Delta q $($\delta q$) 
then separates into two parts: One part includes the contributions from quark 
states only, which corresponds to $\Delta q_{\it q} $($\delta q_{\it q}$). 
Another part includes the contributions of antiquark states, which gives 
$\Delta \bar q $($\delta \bar q)$. Now choosing third($i=3$) component and in terms 
of plane wave solutions, one can write the identical transformations from the 
Dirac spinor to the Pauli spinor for the quark axial and tensor currents as
\begin{equation}
\bar u_{s^{\prime }}(k)\gamma ^3\gamma ^5u_s(k)=(1-\frac{k^2_{\bot }}
{E(E+m)})\chi _{s^{\prime }}^{+}\sigma _3\chi _s,
\end{equation}
\begin{equation}
\bar u_{s^{\prime }}(k)\gamma ^0\gamma ^3\gamma ^5u_s(k)=({\frac mE}+
\frac{k^2_{\bot }}{E(E+m)})\chi _{s^{\prime }}^{+}\sigma _3\chi _s,
\end{equation}
which lead to $M_A=1-\frac{k^2_{\bot}}{E(E+m)}$, $M_T={\frac mE}+
\frac{k^2_{\bot}}{E(E+m)}=1-\frac{k^2_3}{E(E+m)}$. Hence one finds
\begin{equation}
M_A+M_T=1+\frac mE,
\end{equation}
\begin{equation}
1+M_A=2M_T,
\end{equation}
where one assumed $\langle k^2_{\bot} f(\vec k^2) \rangle =2\langle k^2_3 f(\vec k^2)\rangle$.
Eq.(7) was discussed earlier by Close[10]. The relation(10) was also obtained 
by Schmidt and Soffer[6] from the Melosh rotation[11]. Solution of 
combined Eqs.(9) and (10) yields Eq.(5). Combining Eqs.(3)-(5) leads to Eq.(6). 
Note that there is no transformation relation between 
$\Delta \bar q(\delta \bar q)$ and $\Delta q_{NR}(\delta q_{NR})$ because 
$\Delta \bar q(\delta \bar q)$ given through the expansion of the quark field operators
contains antiquark creation and annihilation operators but there is no net 
antiquark in the sNCQM.

It is interesting to remark that $\Delta q_q$ and $\delta q_q$ given by Eqs.(3)-(5)
split up into two parts arising respectively from the valence current quark
contributions, $\Delta q_v$ and $\delta q_v$, and the sea-($q\bar q$)-pair contributions,
$\Delta q_m$ and $\delta q_m$: $\Delta q_q=\Delta q_v +\Delta q_m$, 
$\delta q_q=\delta q_v +\delta q_m$, where
\begin{equation} 
\Delta q_v=\langle \frac 13+\frac{2m_{cu}}{3E}\rangle \Delta q_{NR},\ \ 
\delta q_v=\langle \frac 23+\frac{m_{cu}}{3E}\rangle\delta q_{NR},
\end{equation}
\begin{equation}
\Delta q_m=\langle \frac{2m_{dy}}{3E}\rangle \Delta q_{NR},\ \
\delta q_m=\langle \frac{m_{dy}}{3E}\rangle \delta q_{NR},
\end{equation}
where one has written the constituent quark mass as $m=m_{cu}+m_{dy}$, the 
sum of the current-quark and dynamical masses. To understand the physical 
contents of Eqs.(11)-(12), one considers following cases. (a) $m/E\rightarrow 0$, 
or $|\vec k|\rightarrow \infty$, i.e.the ultra-relativistic limit of quark 
motion. In this limit, Eqs.(3)-(5) 
reduce to $\Delta q_q\rightarrow \frac 13 \Delta q_{NR}$ and
$\delta q_q\rightarrow \frac 23\delta q_{NR}$, which are just
the results for the free valence quarks with zero masses,i.e.the free valence
current quarks. In this limit
the valence current quarks are such free quarks due to the QCD asymptotic
freedom. This shows that the contributions given by Eq.(11) 
come frome the valence current quarks, where the terms 
$\sim \langle \frac{m_{cu}}{E}\rangle$ are 
the correction from current quark mass. (b)$m/E\neq 0$, 
which corresponds to the case of quark motion with finite momentum 
(due to confinement) in the region between the chiral symmetry breaking scale 
($\simeq 1GeV $) and the confinement scale, where the constituent quark 
picture works. One notices that in the matrix elements(1) and (2)
the information on sea-quark is contained in the quark's dynamical
mass parameter if the nucleon wavefunction takes the traditional  
three-quark structure. In fact, in this nonperturbutive region the quarks 
propagate in a ground state filled with $(q\bar q)$ condensates 
generated by spontaneous chiral-symmetry breaking and hence gains the
dynamical mass $m_{dy}$, forming the constituent quarks. So $m_{dy}$ arises from the contribution of 
($q\bar q$)-pair cloud(sea) which surrounds the valence current quark and 
is described by the quark condensate $\langle \bar qq\rangle $[12].
It implies that $\Delta q_m $ and $\delta q_m$ given by Eq.(12) represent 
phenomenologically the polarizations of the ($q\bar q$)-pairs in the 
cloud-sea or sea-($q\bar q$)-pairs briefly. The remainders of polarizations 
are from valence current quarks given by Eq.(11).
It is worth to emphasize that the separation of polarizations 
into valence and sea-($q\bar q$)-pair components follows the general
picture of the constituent quark structure[13]. In the nonrelativistic limit 
$|\vec k|=0$,
$M_A^m\rightarrow 2/3$, $M_T^m\rightarrow 1/3$,
hence $M_A=M_T=1$, which is then consistent.

  Since there is no transformation relation between $\Delta \bar q (\delta \bar q)$ 
and $\Delta q_{NR}(\delta q_{NR})$, the complete expressions for 
$\Delta q$ and $\delta q$ should include antiquark contributions, 
$\Delta \bar q$ and $\delta \bar q$, by definition:
\begin{equation}
\Delta q=\Delta q_v+\Delta q_m+\Delta \bar q=\Delta q_v+\Delta q_{sea},
\end{equation}
\begin{equation}
\delta q=\delta q_v+\delta q_m-\delta \bar q=\delta q_v,
\end{equation}
where $\Delta q_v$, $\Delta q_m$, $\delta q_v$ and $\delta q_m$ are given 
through transformations(11)-(12) by the symmetric three-quark 
structure of the nucleon. $\Delta q_{sea}=\Delta q_m+\Delta \bar q $.
$\Delta \bar q(\delta \bar q)$ may be understood as a result accompanying
the current operator renormalization which leads to the gluon anomaly
contribution[14], or more physically the contribution from 
high Fock states beyond the traditional three-quark structure such as 
$|qqqg\rangle $ which may also lead to the contribution of identifying with the gluon 
anomaly as discussed by Brodsky etc[15] or $|qqq(q\bar q)\rangle $ 
etc in the more realistic nucleon wavefunction. 

According to the above recipe, now one first calculates the proton axial 
charge. By using Eq.(11), one immediately obtains the contributions of the 
valence current quark spins to the axial charge hence to the proton spin
(in the chiral limit with $m_{cu}=0$):
\begin{equation}
\begin{array}{rcl}
 &\Delta u_v=4/9,\ \Delta d_v=-1/9,\ \Delta s_v=0,& \\
 &\Delta \Sigma _v=\Delta u_v+\Delta d_v+\Delta s_v=1/3,&
\end{array}
\end{equation}
which indicates that valence current quark spins carry $1/3$ of the proton
spin. This result is consistent with the asymptotic limit of perturbation QCD 
calculation obtained by Ji etc[9].

To determine the polarizations of antiquarks and quarks in the quark-sea inside
the proton, $\Delta \bar q$, $\Delta q_m$ and $\Delta q_{sea}$, one needs
the complete proton wavefunction of which so far nobody is known. Instead, here one
first gets $\Delta q_{sea}$ from fetting the experimental values, and then
uses transformation(12) to estimate $\Delta q_m$, thus
getting $\Delta \bar q$ by Eq.(13). From recent data on polarized
lepton-nucleon DIS experiments, Ellis and Karliner have shown[4]
\begin{equation}
\Delta u=0.82,\Delta d=-0.44,\Delta s=-0.11,\Delta \Sigma =0.27
\end{equation}
with an estimated error of 0.03 for each flavor's contribution .
Substituting the values given by Eqs.(15) and (16) into Eq.(13), one obtains
the sea polarizations inside the proton:
\begin{equation}
\Delta u_{sea}=0.38,\Delta d_{sea}=-0.33,\Delta s_{sea}=-0.11,\Delta
\Sigma _{sea}=-0.06.
\end{equation}
Furthermore, one estimates $\Delta q_m$ by using Eq.(12), where the
parameter $\left\langle \frac mE\right\rangle $ can be fixed by requiring
$\Delta u_{\it q}-\Delta d_{\it q}=1.257$, which gives $\left\langle \frac
mE\right\rangle =0.631\approx 5/8$. Thus one estimates 
\begin{equation}
\Delta u_m=0.56,\Delta d_m=-0.14,\Delta s_m=0,\Delta \Sigma _m=0.42.
\end{equation} 
Comparing Eq.(17) with Eq.(18), one then has 
\begin{equation}
\Delta \bar u=-0.18,\Delta \bar d=-0.19,\Delta \bar s=-0.11, \Delta \bar
\Sigma =-0.48.  
\end{equation}
These numerical results show the following picture: From the point of view
of the current quark picture, the valence constituent
quark spin contributions given in the traditional symmetric quark model include 
the contributions from the valence current quarks and the sea-($q\bar q$)-pairs:
$\Delta \Sigma _{\it q}=\Delta \Sigma _v+\Delta \Sigma _m=0.33+0.42=0.75$,
where the valence current quark spins contribute $1/3$ of the proton spin.
Antiquarks inside the proton are significantly polarized in the
direction opposite to the proton spin, while the sea-($q\bar q$)-pairs are polarized
with the same direction as the valence current quark polarizations (see Eqs.(15) 
and (18)). As a result, the total contribution of sea polarization,
$\Delta \Sigma _{sea}=\Delta \Sigma _m+\Delta \bar \Sigma $, seems to be
suppressed because of large cancellation between $\Delta\Sigma _m$ and 
$\Delta \bar\Sigma$ as shown by Eq.(17). Eq.(17) implies 
$\Delta \Sigma _{sea}\leq 0$. Combining it with Eq.(15) indicates that 
the total contribution of quark spins to the proton spin  
$\Delta \Sigma =\Delta \Sigma_v+\Delta \Sigma _{sea}$ in the chiral limit with
$m_{cu}=0$ satisfies 
\begin{equation} 
\Delta \Sigma =1/3+\Delta \Sigma _{sea}\leq 1/3 ,
\end{equation}
which indicates the physical origin why $\Delta \Sigma$ extracted from 
the recent global fit to experimental data remains around 0.3 [4].
Including the correction from the current quark mass, which is estimated by
Eq.(11) with $m_{cu}^{(u)}=4 MeV$,
$m_{cu}^{(d)}=8 MeV$ and $m=320 MeV$, one finds 
$\Delta \Sigma_v=0.337$ and $\Delta \Sigma=0.337+\Delta \Sigma_{sea}$, where
$\Delta \Sigma_{sea}=\Delta u_{sea}+\Delta d_{sea}+\Delta s_{sea}+...$.

To understand the origin of the proton spin more clearly, let me
make more complete analysis to the proton spin structure. In QCD,
the proton spin is defined as the expectation value of the
total angular momentum operator $\hat J$ of QCD in the nucleon state, 
where $\hat J$ splits up into quark spin and orbital and gluon angular
momentum components[9,16]. The quark model might be imagined as the low-energy   
limit of QCD, where the gluons are integrated out and the quarks become the
effective degrees of freedom of QCD. Thus, in the spirit of the quark model,
$\hat J$ separates into quark spin and orbital parts which have same form as that
in QCD[9,16]. The third component of $\hat J$ reads 
\begin{equation}
\hat J^3=\int d^3x[\frac 12\bar q\gamma ^3\gamma ^5q+i\bar q\gamma
^0(x^1\partial ^2-x^2\partial ^1)q].
\end{equation}
By the same procedure as the derivation of Eqs.(3)-(5), one obtains that the
quark orbital contribution is related to the sNCQM's spin 
contribution (for each quark flavor) by 
\begin{equation}
L_3^{(q)}=\langle M_L\rangle \Delta q_{NR}
\end{equation}
with $M_L=\frac{\vec k^2}{3E(E+m)}=\frac 13-\frac m{3E}$. Combining Eqs.(3),
(5), (22) with(21), one then obtain a spin sum rule
\begin{equation}
\frac 12=\frac 12\Delta \Sigma _v+\frac 12\Delta \Sigma _m+L_{zq},
\end{equation}
where $L_{zq}=L_3^{(u)}+L_3^{(d)}=\frac 13-\langle \frac m{3E} \rangle $. 
One may write $L_{zq}=L_{zv}+L_{zm}$, where $L_{zv}=1/3$ and
$L_{zm}=-\langle\frac m{3E}\rangle$, that is, the quark orbtial contribution
separates into valence current quark plus sea-($q\bar q$)-pair parts. It is intresting
that $\frac 12\Delta \Sigma_m +L_{zm}=0$, i.e. the orbital and spin 
contributions of sea-($q\bar q$)-pairs cancel each other. As a result, Eq.(23) in fact becomes 
\begin{equation}
\frac 12=\frac 12\Delta \Sigma_v +L_{zv},\ \ \Delta \Sigma_v=L_{zv}=\frac 13,
\end{equation}
which clearly indicates that $\Delta \Sigma_v$ indeed is the valence current
quark spin contribution to the proton spin and that the basic origins of the 
proton spin in the symmetric (valence) quark model are valence current quark spin
and orbital contributions. The fractions of the proton spin are re-shared due 
to the excitations of the sea-($q\bar q$)-pairs, which is expressed by Eq.(23) 
where the spin of the proton is shared by the valence current quark spin, 
spin of sea-($q\bar q$)-pair 
and quark orbital contributions. Their values are ${\frac 1 6}$, 0.21, and 
0.12, respectively, at scale of 1$GeV$. Only in the nonrelativistic limit 
$E=m$, $L_{zq}$ disappears and hence Eq.(23) reduces to 
$\Delta \Sigma_{NR}=1$, which of course is not a realistic case because 
$L_{zq}\not=0$ as long as quarks have non-zero momentum as shown by 
Eq.(22). One may ask why one did not find $\Delta \Sigma _q =\Delta \Sigma_v+
\Delta \Sigma_m =0.75$ from polarized DIS? The reason is that in 
polarized DIS what one probes is the total contribution of quark spins, 
$\Delta \Sigma =\Delta \Sigma _v+\Delta \Sigma _m+\Delta \bar \Sigma $. Its
value $\leq 1/3$. This result clearly indicates the physical origin of the
discrepancy between the traditional quark model expectation $\Delta \Sigma _q$ 
and the measured value $\Delta \Sigma $ in polarized DIS: Antiquarks are 
significantly polarized inside the proton, which has not been counted in the traditional 
quark model of the proton. Besides this, the sNCQM has neglected the quark 
motion effects which lead to the reduction of $\Delta \Sigma_{NR}$ and the 
presence of quark orbital angular momentum.
 
 Considering the contents of $\Delta \Sigma $ measured in polarized DIS, one 
now rewrites Eq.(23) as
\begin{equation} 
\frac 12=\frac 12\Delta \Sigma +L_{zq}+J_{\bar q} .  
\end{equation} 
where $J_{\bar q}=-\frac 12\Delta \bar \Sigma$.
Eq.(25) can be compared with the spin sum rule in QCD[16]. In fact, as mentioned 
before, $\Delta \bar \Sigma$ may be understood as a result due to existing
the gluon anomaly contribution. Thus Eq.(25) means that the fractions of the 
proton spin are further re-shared due to the existence of the gluon degrees
of freedom. Numerically one has $\frac 12\Delta \Sigma =0.14$, $L_{zq} \simeq 0.12$ 
and $J_{\bar q}=\frac 12(-\Delta \bar \Sigma)\simeq 0.24$. Thus $J_{q}= 
\frac 12\Delta \Sigma +L_z \simeq 0.26 $, and $J_{q}\leq 0.29 $ by Eq.(20). This estimate
shows that quarks (their spin plus orbital contributions) carry about one half 
of the proton spin at scale of 1$GeV$. The remainder of the proton spin is
carried by gluons,
which may be estimated in number by the absolute value of the 
antiquark contribution to the proton spin in the quark model.

 Now one turns to the tensor charge of the proton. Since the proton
tensor charge measures the contributions from valence current quarks only, one can
naturally obtain it by using Eq.(11). The result is(in the chiral limit with
$m_{cu}=0$):
\begin{equation}
\delta u=\frac 89,\ \ \delta d=-\frac 29,\ \
g_T^{(v)}=\frac{10}9,\ \ g_T^{(s)}=\frac 23,
\end{equation}
where $g_T^{(v)}=\delta u-\delta d$, and $g_T^{(s)}=\delta u+\delta d$ are
the proton's isovector and isoscalar tensor charges, respectively. One may
argue the reliability of the result given by Eq.(26). Indeed, by using
Eqs.(4) and (5) one obtains the tensor charge of the proton in the symmetric 
quark model: 
$\delta u_{\it q}=\frac 89+\frac 49\times 0.631=1.17$, $\delta d_{\it q}
=-\frac 29-\frac 19\times 0.631=-0.29$, 
which are exactly values
obtained in the MIT bag model[5] and in the Melosh rotation approach[6]. 
However, one should recall that these values include the contributions from 
both valence current quarks and the sea-($q\bar q$)-pairs:
$\delta q_{\it q}=\delta q_v+\delta q_m$. As pointed out before, the
high Fock states in the more realistic nucleon wavefunction may contribute 
$\delta \bar q$, thus leading to $\delta q_m-\delta \bar q=0$ because sea 
quarks do not contribute to the tensor charge. As a result, one has 
$\delta q=\delta q_v$. Therefore, the result given by Eq.(26) expresses the 
proton tensor charge in the chiral limit at the scale $\mu _0 \simeq 1GeV$. 
The correction from current quark mass, estimated by Eq.(11), for $\delta u$
is 0.004 and for $\delta d$ is $-$0.002. The tensor charge at 
any scale $\mu$ can be obtained through the
 evolution equation $\delta q(\mu ^2)=\delta q(\mu _0^2)[\alpha _s(\mu
^2)/\alpha _s(\mu _0^2)]^{\frac 4{33-2n_f}}$, where $n_f$ is the number of
quark flavors. It is expected that the future experiments will test the
present predicition.

\acknowledgements
\medskip
 I thank X.Ji for helpful comments. This work is supported in part by the National Natural Science Foundation
of China and the Nuclear Science Foundation of China.


\begin{references}
\bibitem{8} Another important issue is how to decompose the nucleon spin into 
quark and gluon contributions and to calculate them in QCD, which has been 
nicely discussed recently by Ji etc[9].
\end{references}
\end{document}